\newcommand{\simgt}{\lower.5ex\hbox{$\; \buildrel > \over \sim \;$}}
\newcommand{\simlt}{\lower.5ex\hbox{$\; \buildrel < \over \sim \;$}}
\newcommand{\solM}{\mathrm{M_{\odot}}}
\documentclass{ws-ijmpd}
\usepackage[usenames]{color}
\begin{document}

\markboth{T.~Narikawa, R.~Kimura, T.~Yano, K.~Yamamoto}
{Halo models in modified gravity theories 
with self-accelerated expansion}

%
\catchline{}{}{}{}{}
%
\title{HALO MODELS IN MODIFIED GRAVITY THEORIES WITH \\
SELF-ACCELERATED EXPANSION}

\author{TATSUYA~NARIKAWA}
\address{Department of Physical Science, Hiroshima University,
Higashi-Hiroshima 739-8526,~Japan\\
narikawa@theo.phys.sci.hiroshima-u.ac.jp}

\author{RAMPEI~KIMURA} 
\address{rampei@theo.phys.sci.hiroshima-u.ac.jp}

\author{TATSUNOSUKE~YANO} 
\address{yano@theo.phys.sci.hiroshima-u.ac.jp}

\author{KAZUHIRO~YAMAMOTO}
\address{kazuhiro@hiroshima-u.ac.jp}

\maketitle


\begin{abstract}
We investigate the structure of halos in the sDGP (self-accelerating branch of 
the Dvali-Gavadadze-Porrati braneworld gravity) model and the galileon 
modified gravity model on the basis of the static and spherically 
symmetric solutions of the collisionless Boltzmann equation, 
which reduce to the singular isothermal sphere model and the King model 
in the limit of Newtonian gravity. 
The common feature of these halos is that the density of a halo 
in the outer region is larger (smaller) in the sDGP (galileon) 
model, respectively, in comparison with Newtonian gravity. 
This comes from the suppression (enhancement) of the effective 
gravity at large distance in the sDGP (galileon) model, respectively. 
However, the difference between these modified gravity models
and Newtonian gravity only appears outside the halo due 
to the Vainshtein mechanism, which makes it difficult to 
distinguish between them. We also discuss the case in which 
the halo density profile is fixed independently of the gravity 
model for comparison between our results and previous work. 
\end{abstract}

\def\bfx{{\bf x}}
\def\bfk{{\bf k}}
\def\mpl2{{M_{\rm Pl}^2}}
\def\hmpc{h^{-1}{\rm Mpc}}

\section{Introduction}
Motivated from the discovery of the cosmic accelerated expansion of 
the universe\cite{Riess,Perlmutter}, it is becoming important to 
test the nature of gravity on the cosmological scales. This is because 
the nature of gravity might be deeply rooted to the cosmic accelerated 
expansion (e.g., Refs.~\refcite{Maartens,Jain10,Tsujikawa}). 
In general, it is very challenging to construct a modified gravity 
(MG) theory that is consistent with observations.
In comparison with the varieties of viable dark energy models to
explain the accelerated expansion of the universe, 
many modified gravity models are ruled out by cosmological observations
(e.g., Refs.~\refcite{Maartens,Jain10,Tsujikawa,DE06,AmeTsujiDE}). 

{}For example, one of the most popular modified gravity models is the 
Dvali-Gavadadze-Porrati (DGP) model\cite{DGP,DGP2}.
The DGP model is described in the context of the braneworld scenario,
which consists of a 3+1-dimensional brane embedded in a 5-dimensional 
bulk. This model has an interesting phenomenology and yields two 
branches of the Friedmann equation, 
which are a self-accelerating branch (sDGP) 
and a normal branch (nDGP).
In the sDGP model, the expansion of the universe self-accelerates 
at late times without a cosmological constant nor spatial curvature, 
while the nDGP model needs to add a stress-energy component with 
negative pressure on the brane to be consistent with cosmological 
observations. 
Unfortunately, the sDGP model inevitably give rise to 
a ghost mode\cite{KoyamaMaartens,Nicolis04,Gorbunov06}.
Moreover, the sDGP model is disfavored by 
the cosmological observations\cite{Fairbairn,Maartens06,Song07}.

Inspired by the decoupling limit of the DGP model, the galileon gravity 
theory has been studied as a possible alternative to large 
distance modification of gravity (e.g., 
Refs.~\refcite{DGP3,Luty03,GALMG,CG,GC,SAUGC,ELCP,CEGH,Deffayet,GInflation,DeFelice10,CCGF,OCG,Kimura}).
This theory introduces a scalar field with the kinetic term 
with the ''wrong'' sign and the self-interaction term 
$(\partial\phi)^2\square\phi$, which is invariant under 
the Galilean shift symmetry $\partial_\mu\phi\rightarrow\partial_\mu\phi+b_\mu$
in the Minkowski space-time, which keeps equation of motion at the
second order differential equation. 
Although the Lagrangian no longer satisfies the Galilean shift symmetry in 
a curved spacetime, the field equation of the galileon field remains 
a second order differential equation, and 
the galileon field admits the self-accelerating solution 
in a FRW universe without a ghost instability.

The other notable feature of these modified gravity models is 
the Vainshtein mechanism, 
which allows the modification of gravity to recover 
general relativity around a high density region\cite{Vainshtein}.
Thanks to the Vainshtein mechanism, these models evade the solar 
system constraint. 
It is worth examining whether or not the effect of modification 
of gravity on the nonlinear scales and the structure of 
the halo of galaxy or galaxy cluster might be useful as a clue 
to test these modified gravity theory. 
Recently, several researchers have investigated such a possibility of testing 
these modified gravity models on the scales of galaxy or galaxy 
cluster\cite{dynamicalmass,Wyman,Konno,Gergely,Burikham}. 
In the present paper, we focus our investigation on the structure 
of a halo in the sDGP model and the galileon model by constructing
numerical solutions that correspond to the singular isothermal 
sphere (SIS) and the King model in the limit of Newtonian gravity.

The structure of this paper is as follows.
In Sec.~\ref{MGmodel}, we first give a brief review 
of the sDGP model and the galileon model. Then, we 
give basic formulas for our halo models in the sDGP 
model and the galileon model.  
In Sec.~\ref{numericalresult}, our numerical
results of the halo models are demonstrated. 
In Sec.~\ref{discussion}, we investigate the case with a different 
approach for the halo modeling, in which the halo density profile 
is fixed independently of the gravity model for comparison between 
our results and previous work\cite{dynamicalmass}. 
Sec.~\ref{conclusion} is devoted to summary and conclusions.
Throughout this paper, we use units in which the speed of light
equals unity, and we follow the metric convention $(-,+,+,+)$.
We use the reduce Planck mass $M_{\rm Pl}$, which is defined by
$M_{\rm Pl}=1/\sqrt{8\pi G}$ with Newton's gravitational constant $G$.
We adopt the Hubble constant $H_0=100~h{\rm km/s/Mpc}$ with $h=0.7$ and 
the matter density parameter at present $\Omega_0=0.28$.

\section{Basic Formulas}
\label{MGmodel}
\subsection{Modified gravity models}

The Dvali-Gabadadze-Porrati model consists of 
a spatially three-dimensional brane in a 
4+1 dimensional (5D) Minkowski bulk. 
The action is\cite{DGP,DGP2} 
\begin{eqnarray}
S={M_5^3\over2}\int d^5x\sqrt{-g_5}R_5
+\int d^4x\sqrt{-g}\biggl({\mpl2\over2}R+{\cal L}_{\rm m}\biggr),
\end{eqnarray}
where $M_5$ ($M_{\rm Pl}$) is the Planck mass in the 5 (4) dimensional 
spacetime, $g_5$ ($g$) and $R_5$ ($R$) are the determinant and the 
Ricci scalar of the 5 (4) dimensional metric, respectively, and 
${\cal L}_{\rm m}$ stands for the matter Lagrangian.
The two Planck masses $M_5$ and $M_{\rm Pl}$ can be related via  
length scale, the crossover scale $r_c \equiv \mpl2/2M_5^3$, 
which must be fine-tuned to be the present-day horizon scales 
in order to modify gravity only at late times.
In this model, the Friedmann equation has two branches. 
One corresponds to the self-accelerating solution 
called sDGP, which contains a ghost-like instability. 
Although the other solution called nDGP
does not suffer from a ghost instability, 
the cosmological constant is needed to drive the cosmic acceleration.
In the present paper,
we consider the self-accelerating branch of the DGP model.
During matter domination and beyond, 
the modified Friedmann equation in the sDGP model is\cite{DGP3}
\begin{eqnarray}
{H(a)\over H_0}={1-\Omega_0\over 2}+\sqrt{{\Omega_0\over a^{3}}+{(1-\Omega_0)^2
\over 4}}~,
\end{eqnarray}
where $a$ is the scale factor, and the matter density parameter 
is related to the crossover scale by $r_c=1/(1-\Omega_0)H_0$. 
Note that the modification of the law of gravity in the nDGP model
is qualitatively similar to that in the galileon model. 
Therefore, our results in the galileon model
can be applied to the nDGP model.

On the other hand, the galileon model is characterized 
by a scalar field with the self-interaction
whose Lagrangian is invariant under the Galilean shift symmetry
in the Minkowski spacetime\cite{GALMG,CG}. 
We consider the galileon model in a curved spacetime, which is 
minimally coupled to gravity, with the action\cite{Deffayet,GInflation},
\begin{eqnarray}
S=\int d^4x\sqrt{-g}\left[{{\mpl2\over 2}R}+K(X)
-G(X)\square\phi+{\cal L}_{\rm m}\right],
\end{eqnarray}
where $R$ is the Ricci scalar,
$X=-g^{\mu\nu}\nabla_\mu\phi\nabla_\nu\phi/2$, 
~$\square\phi=g^{\mu\nu}\nabla_\mu\nabla_\nu\phi$,  
${\cal L}_{\rm m}$ is the matter Lagrangian, and
$K(X)$ and $G(X)$ is an arbitrary function of $X$. 
{}For simplicity, we consider the following functions, 
$K(X)=-X$ and $G(X)={(r_c^2/{M_{\rm Pl}})}X$,
where $r_c$ is the model parameter.
This model 
admits a late-time de-Sitter attractor in a flat FRW 
universe.
The solution along this attractor
can remarkably simplify the modified Friedmann equation, and 
the background evolution during the matter dominated era can be
regarded as the Einstein de-Sitter universe. Besides, this model 
is not plagued by ghost instability in contrast to the sDGP model.
In the present paper, we consider the attractor solution. 
Then, the modified Friedmann equation on the attractor can be 
written as
\begin{eqnarray}
\left({H(a)\over H_0}\right)^2={1\over2}
\biggl[\Omega_{0}a^{-3}+\sqrt{\left(\Omega_{0}a^{-3}\right)^2+4(1-\Omega_{0})}\biggr].
\end{eqnarray}
The parameter $r_c$ is related with the cosmological 
parameters through the relation, $r_c=1/(54(1-\Omega_0))^{1/4}H_0^{-1}$.

Extended models of the galileon model have been 
proposed\cite{Deffayet,Kimura}. In Ref.~\refcite{Kimura}, the model
with $G(X)=M_{\rm Pl}{(r_c^2/{M^2_{\rm Pl}})^n}X^n$ is considered, 
and it is demonstrated that this model reduces to the 
cosmological constant model for $n$ equal to infinity. 
Observational constraints on this model is also investigated, 
focusing on the constraints from the Ia supernovae (SN) observations 
and the cosmic microwave background (CMB) distance observation, 
as well as from the large scale structure of the luminous
red galaxies in the sloan digital sky survey data release 7. 
The model with $n=1$ is disfavored by the constraints 
from the SN and CMB observations, however, the model with 
higher $n$ can be consistent with the observations. 
In the present paper, we consider the halo models of the
galileon model with $n=1$, however, the result is almost 
same as those with higher $n$ at a quantitative level. 

\subsection{Perturbation equations in the static approximation}
In this subsection, we summarize basic perturbation equations for 
gravity and the brane bending mode in the sDGP model and the galileon field on 
small scales, assuming the spherically symmetric and static system. 
{}For convenience, we choose the Newtonian gauge, which is given by
\begin{eqnarray}
 ds^{2}=-(1+2\Psi)dt^{2}+a^2(t)
(1+2\Phi)\delta_{ij}dx^{i}dx^{j},
\end{eqnarray}
where $a(t)$ is the scale factor, but we
set $a(t)=1$ in the expressions below, for simplicity.
{}Following Refs.~\refcite{KoyamaMaartens} and \refcite{Koyama07} 
for the DGP model, and Refs.~\refcite{SAUGC} and \refcite{Kimura} for 
the galileon model, within the 
subhorizon scales with the quasi-static approximation,  
we have the following perturbed equations,
\begin{eqnarray}
&&\triangle\Phi=
-{4\pi G}\rho +\xi\triangle \varphi \label{pE}, \\
&&\Phi+\Psi=-\alpha\varphi,
\label{pE2}
\end{eqnarray}
and 
\begin{eqnarray}
\triangle\varphi+\lambda^2(\varphi_{,ij}\varphi^{,ij}-
(\triangle\varphi)^2)=-4\pi G\zeta\rho,
\label{gE}
\end{eqnarray}
where $\varphi(x)$ denotes the brane
bending mode and the perturbation of 
the galileon field defined by 
$\phi(t,x)=\phi(t)(1+\varphi(x))$,
the Laplace operator $\triangle$ represents the differentiation 
with respect to physical space-coordinates, 
$\rho$ is the matter density.
Here, $\alpha$, $\xi$, $\zeta$, $\lambda^2$ and $\beta$ are determined by the 
background evolution as follows: 
\begin{eqnarray}
&&
\alpha=-1,
~~~~~
\xi = {1\over2},
~~~~~
\zeta= -{2\over3\beta}, 
~~~~~
\lambda^2= -{r_c^2\over3\beta},
~~~~~
\beta= 1-2Hr_c\left(1+{\dot{H}\over3H^2}\right),
\nonumber\\
\end{eqnarray}
for the sDGP model\cite{KoyamaMaartens,Koyama07}, and 
\begin{eqnarray}
&&
\alpha=0,
~~~~~
\xi = {4\pi G G_X}\dot{\phi}^{2}\phi,
~~~~~
\zeta= {G_{X}\dot{\phi}^{2}\over\beta\phi}, 
~~~~~
\lambda^2= {G_{X}\phi \over \beta},
~~~~~
\nonumber\\
&&\beta= -1+2G_{X}(\ddot{\phi}+2H\dot{\phi})-4\pi G G_{X}^{2}{\dot\phi}^{4}, 
\end{eqnarray}
for the galileon model, 
where we defined $G_X=\partial G(X)/\partial X$.
For the galileon model, it is useful to rewrite the combinations 
$\xi\zeta$ and $\lambda^2\zeta$ in terms of the matter density parameter 
$\Omega_m=\rho_m(a)/3\mpl2 H^2(a)$, 
\begin{eqnarray}
\xi\zeta={(1-\Omega_m)(2-\Omega_m)\over \Omega_m(5-\Omega_m)},
~~~~~
\lambda^2\zeta=\left({2-\Omega_m\over H\Omega_m(5-\Omega_m)}\right)^2,
\end{eqnarray}
which is derived along the attractor solution (see e.g., Ref.~\refcite{Kimura}).

In the spherically symmetric case, Eqs.~(\ref{pE}), (\ref{pE2}) and (\ref{gE}) 
reduce to
\begin{eqnarray}
&&{d\Psi\over dr}={GM(r) \over r^{2}} - (\alpha+\xi){d\varphi \over dr}, 
\label{eq:Poisson}
\\
&&{d\varphi\over dr}= {r \over 4\lambda^{2}}
\left(1-\sqrt{1+{8G\lambda^{2}\zeta M(r)\over r^{3}}}\right),
\label{eq:G_Field}
\end{eqnarray}
where we define the enclosed mass $M(r)=4\pi \int_0^rdr'r'^2\rho(r')$.
The influence of $\varphi$ is determined by the second term 
in the square root in Eq.~(\ref{eq:G_Field}),
which is characterized by the so-called Vainshtein radius, 
which we defined by 
\begin{eqnarray}
&&r_V= \left[8G\lambda^{2}\zeta M(r_V)\right]^{1/3}.
\label{eq:V_Radius}
\end{eqnarray}
On scales smaller than the Vainshtein radius $r\ll r_V$, 
Eq.~(\ref{eq:G_Field}) gives
$d\varphi / dr = -\sqrt{G\zeta M(r) / 2\lambda^{2}r} \ll GM(r) / r^{2}$. 
Therefore, the law of gravity in the Vainshtein limit reduces to 
Newtonian gravity,
\begin{eqnarray}
{d\Psi\over dr}\simeq {GM(r) \over r^{2}}.
\label{eq:Psismaller}
\end{eqnarray}
On scales larger than the Vainshtein radius $r\gg r_V$,
Eq.~(\ref{eq:G_Field}) gives
$d\varphi / dr = -G\zeta M(r) / r^{2}$. Thus,
the modification of gravity at large distance becomes
\begin{eqnarray}
{d\Psi\over dr}\simeq {G_{\rm eff}M(r) \over r^{2}},
\label{eq:Psilarger}
\end{eqnarray}
where $G_{\rm eff}=G(1+(\alpha+\xi)\zeta)$ is the effective 
gravitational constant in the linearized limit.
$G_{\rm eff}/G$ is always less than unity in the sDGP model, 
while $G_{\rm eff}/G$ is always larger than unity 
in the galileon model (also in the nDGP model). 

\subsection{Solution of Boltzmann equation}
Next, let us consider the equation that the matter component obeys.
Since the distribution of the matter is determined through the 
gravitational potential $\Psi$, the matter follows the usual 
collisionless Boltzmann equation. 
The static solution of the Boltzmann equation can be obtained 
by the Jean's theorem\cite{Binney}. 
We neglect the effect of time-dependent quantities in the 
modified Poisson equation (\ref{pE}) and the field equation 
(\ref{gE}), which might leads to the time-evolution of the 
effective gravitational constant. 
However, these time-dependent quantities evolve on
cosmological time scales. Therefore, we can assume this effect can 
be neglected when considering the static solution of a halo.

In the present paper, we consider the singular isothermal sphere 
(SIS) model and the King model (see e.g., Ref.~\refcite{Binney}). 
The distribution function is given by 
\begin{eqnarray}
    f({\cal E})= 
\displaystyle{{\rho_{1}\over(2\pi\sigma^{2})^{3/2}}e^{{\cal E}/\sigma^{2}}},
\end{eqnarray}
for the SIS model, and
\begin{eqnarray}
    f({\cal E})=\left\{ 
\begin{array}{cr}
\displaystyle{{\rho_{1}\over(2\pi\sigma^{2})^{3/2}}
(e^{{\cal E}/\sigma^{2}} -1)} ~~&{\cal E}>0, \\
\displaystyle{~~~}\\
\displaystyle{0} ~~&{\cal E}\leq 0,\\
\end{array} \right.
\end{eqnarray}
for the King model, respectively,
where $\rho_1$ is a constant, ${\cal E}= -\Psi - s^{2} / 2$, $s$ is 
velocity, and $\sigma$ is velocity dispersion.
By integrating the distribution function over all velocities, 
we have the formula for the density, 
\begin{eqnarray}
\rho=\rho_1 e^{-\Psi/\sigma^2},
\label{eq:DF_SIS}
\end{eqnarray}
for the SIS model, and
\begin{eqnarray}
\rho=\rho_1\biggl[ e^{-\Psi/\sigma^2}{\rm erf}\biggl( \sqrt{-\Psi\over \sigma^2}
\biggr)
-\sqrt{-4\Psi\over \pi \sigma^2}\biggl(1-{2\Psi\over 3\sigma^2}\biggr)\biggr],
\label{eq:DF_King}
\end{eqnarray}
for the King model, respectively, where ${\rm erf}(x)$ is the error 
function defined by ${\rm erf}(x)=(2/\sqrt{\pi})\int_0^xe^{-t^2}dt$. 

The basic equations for a halo are 
(\ref{eq:Poisson}), (\ref{eq:G_Field}) with  (\ref{eq:DF_SIS})
for the SIS model, but with (\ref{eq:DF_King}) for the King model. 
These equations are numerically solved in the next section. 
In the case of Newtonian gravity, the solutions are well 
known\cite{Binney}. Especially, the solution of the SIS model is 
written in the simple analytic form, $\rho(r)={\sigma^{2}/ 2\pi Gr^{2}}$.

\begin{figure}[t]
  \begin{tabular}{cc}
    \begin{minipage}{0.5\textwidth}
      \begin{center}
\includegraphics[width=6.2cm,height=6.2cm,clip]{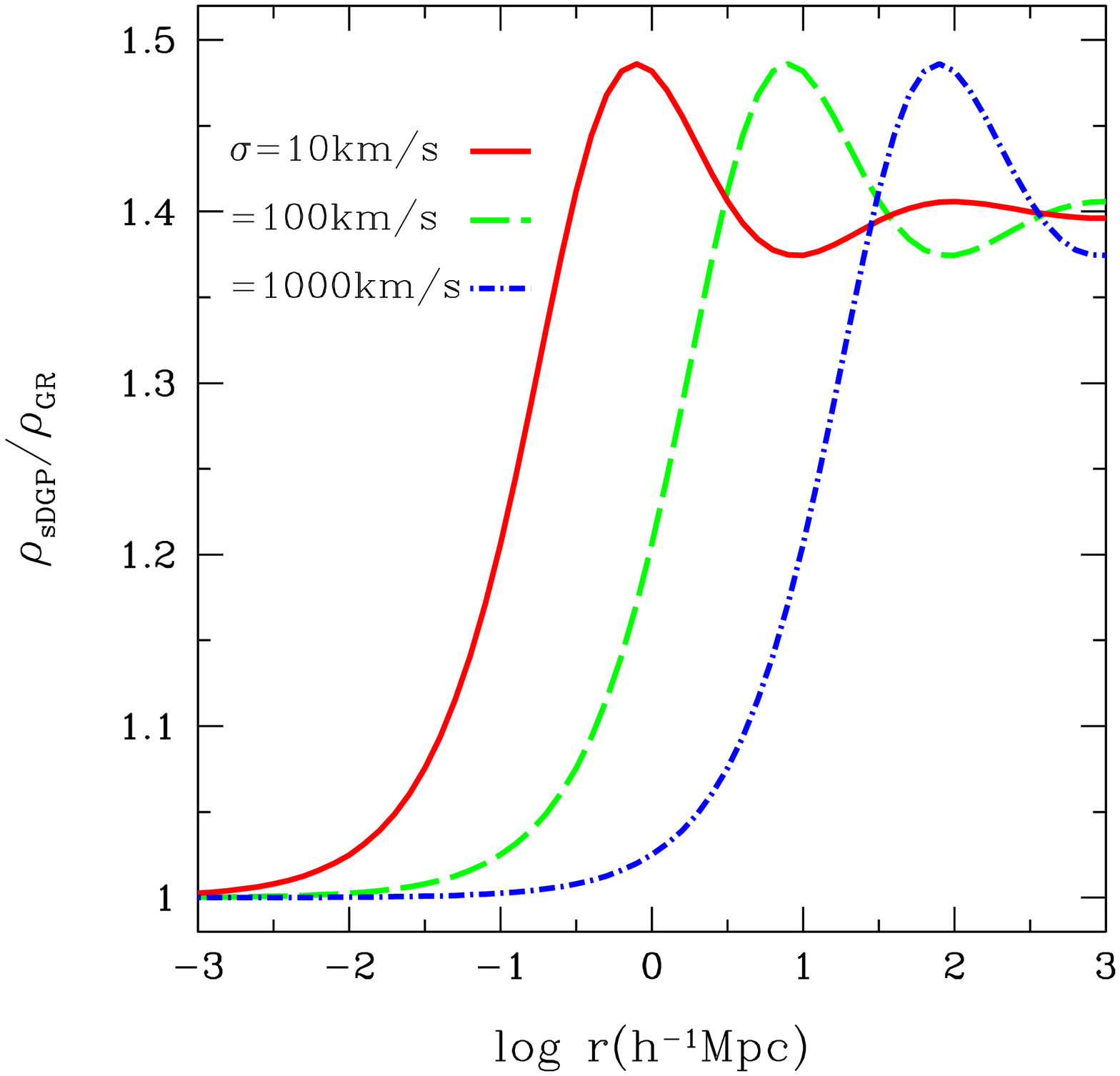}
      \end{center}
    \end{minipage}
    \begin{minipage}{0.5\textwidth}
      \begin{center}
\includegraphics[width=6.2cm,height=6.2cm,clip]{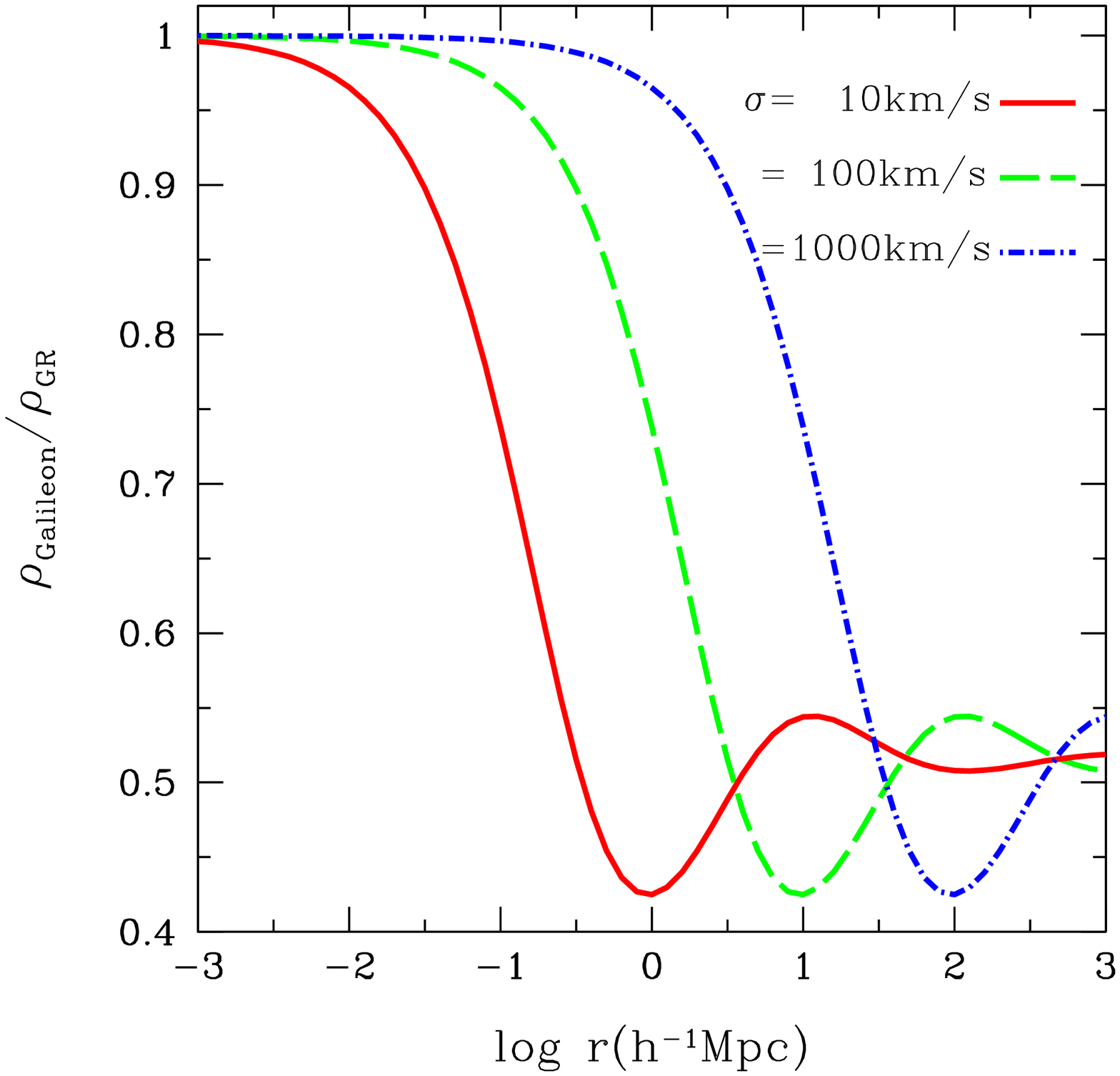}
      \end{center}
    \end{minipage}
  \end{tabular}
\caption{
The left panel is the ratio of the density of the SIS halo
in the sDGP model to that of Newtonian gravity,
as a function of radius $r$. Here, we adopted the velocity 
dispersion $\sigma=10{\rm km/s}$ (solid curve), 
$\sigma=100{\rm km/s}$ (dashed curve), 
and $\sigma=1000{\rm km/s}$ (dot-dashed curve), respectively.
The right panel is the same as the left panel, but in 
the galileon model.
}
\label{fig:SIS_dens}
\end{figure}
\section{Numerical results}
\label{numericalresult}
\subsection{Singular isothermal sphere (SIS) model}

Now, we are ready to solve Eqs. (\ref{eq:Poisson}), (\ref{eq:G_Field}),
and (\ref{eq:DF_SIS}) to find the density profile of the SIS halo 
in the sDGP model and the galileon model. 
Figure \ref{fig:SIS_dens} shows the density of the SIS halo divided by 
that of Newtonian gravity $\rho_{\rm GR}(r)$, as a function 
of the radius $r$, where 
$\rho_{\rm GR}(r)=\sigma^{2}/2\pi Gr^{2}$.
The left panel is the sDGP model, while the right 
panel is the galileon model. Each curve corresponds to the velocity 
dispersion $\sigma=10{\rm km/s}$ (solid curve), 
$\sigma=100{\rm km/s}$ (dashed curve), 
and $\sigma=1000{\rm km/s}$ (dot-dashed curve), respectively.
Here, we used the present values of the Hubble parameter and
the matter density parameter in evaluating $\beta$, $\zeta$, $\lambda^2$, 
and $\xi$.

The density ratio, $\rho_{\rm MG}/\rho_{\rm GR}$, is almost unity at 
small radii $r$ in both the models.
This means that the central region is the same as that of Newtonian gravity. 
This is due to the Vainshtein mechanism. 
The difference appears as the radius $r$ increases. 
We note that the density ratio finally approaches constant 
values as $r$ becomes very large. 
Thus, there are two asymptotic regions, i.e., 
the Vainshtein limit $r\ll r_V$ and the linearized limit $r\gg r_V$. 

In the Vainshtein limit $r\ll r_V$, from Eqs. (\ref{eq:Poisson}), 
(\ref{eq:G_Field}) and (\ref{eq:DF_SIS}), we obtain the approximate
density profile, 
\begin{eqnarray}
\rho_{\rm MG}(r) \simeq \rho_{\rm GR}(r)
{\rm exp}\left(- {(\alpha+\xi)\zeta \over 
\sqrt{\lambda^{2}\zeta}\sigma}r\right).
\label{eq:rhoGali_0}
\end{eqnarray}
Thus, the effect of modification of gravity is represented 
by the factor $(\alpha+\xi)\zeta/\sqrt{\lambda^{2}\zeta}\sigma$. 
On the other hand, in the linearized limit $r \gg r_{\rm V}$, 
Eqs. (\ref{eq:Psilarger}) and (\ref{eq:DF_SIS}) give the 
approximate solution
\begin{eqnarray}
\rho_{\rm MG}(r\gg r_V) \simeq {G\over G_{\rm eff}}\rho_{\rm GR}(r).
\label{eq:rhoGali_lr}
\end{eqnarray}
Thus, the density is enhanced (decreased) 
in the sDGP (galileon) model at large radii $r$, reflecting
the behavior of $G/G_{\rm eff}=1/(1+(\alpha+\xi)\zeta)$. 

Let us estimate the Vainshtein radius $r_V$, which can be
easily done by substituting the relation $M=2\sigma^2r_V/G$, 
extrapolated from Newtonian gravity, into Eq.~(\ref{eq:V_Radius}).
Then, we have the approximate formula of the Vainshtein radius,
\begin{eqnarray}
r_V\simeq4\sqrt{\lambda^{2}\zeta}\sigma.
\label{eq:app_r_v}
\end{eqnarray}
This gives a good approximate expression. However, 
the exact solution of Eq.~(\ref{eq:V_Radius}) gives a
slightly different value from Eq.~(\ref{eq:app_r_v}),
due to the underestimation (overestimation) of 
the mass of the halo in the sDGP (galileon) model
at $r\sim r_V$. 
Our numerical analysis of Eq. (\ref{eq:V_Radius}) shows that
the following formula works,
\begin{eqnarray}
r_V\simeq2.5\left({\sigma\over 100{\rm km/s}}\right)\hmpc,
\end{eqnarray}
for the sDGP model, and 
\begin{eqnarray}
r_V\simeq4.2\left({\sigma\over 100{\rm km/s}}\right)\hmpc,
\end{eqnarray}
for the galileon model, respectively.

\begin{figure}[t]
  \begin{tabular}{cc}
    \begin{minipage}{0.5\textwidth}
      \begin{center}
\includegraphics[width=6.2cm,height=6.2cm,clip]{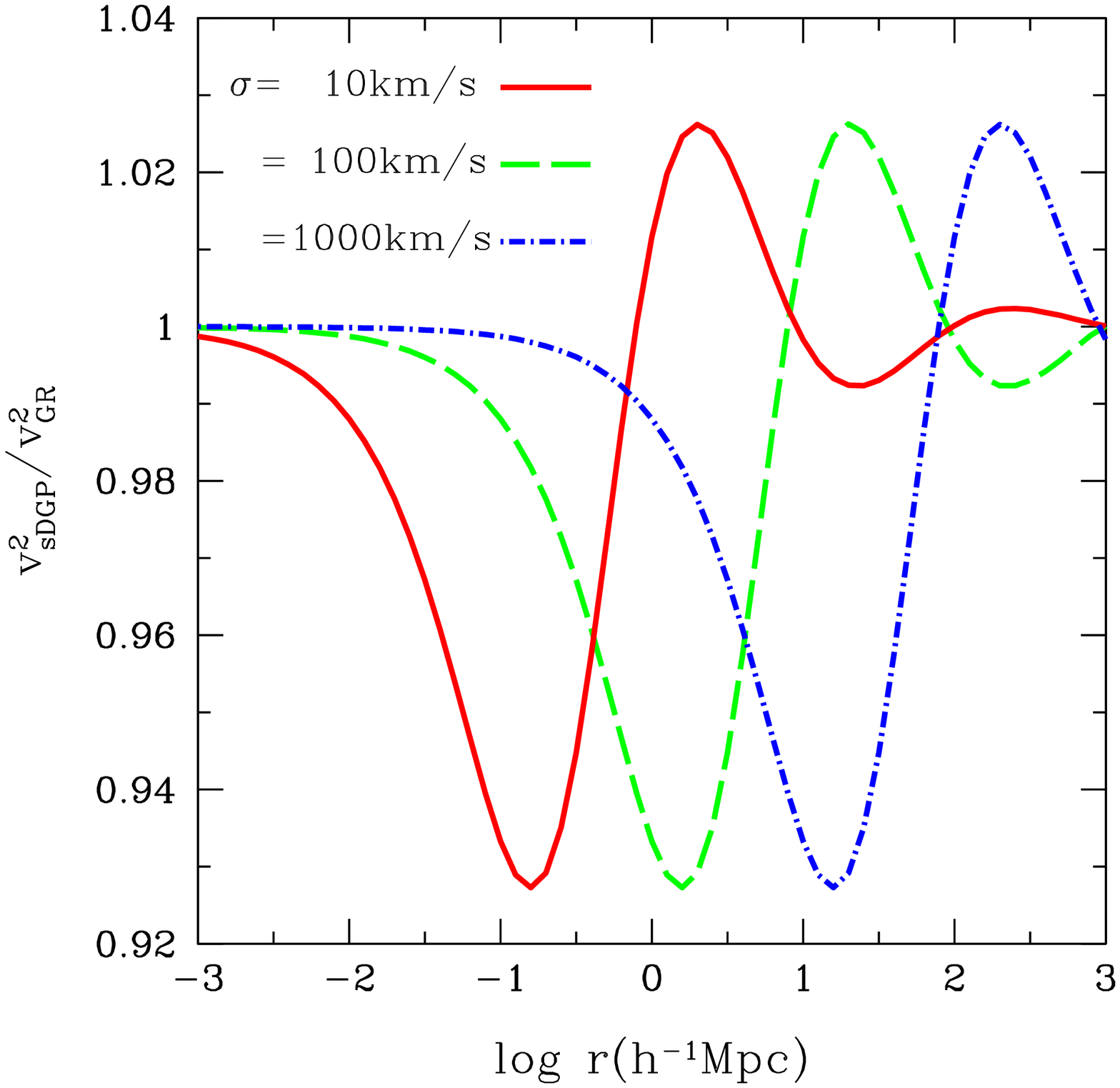}
      \end{center}
    \end{minipage}
    \begin{minipage}{0.5\textwidth}
      \begin{center}
\includegraphics[width=6.2cm,height=6.2cm,clip]{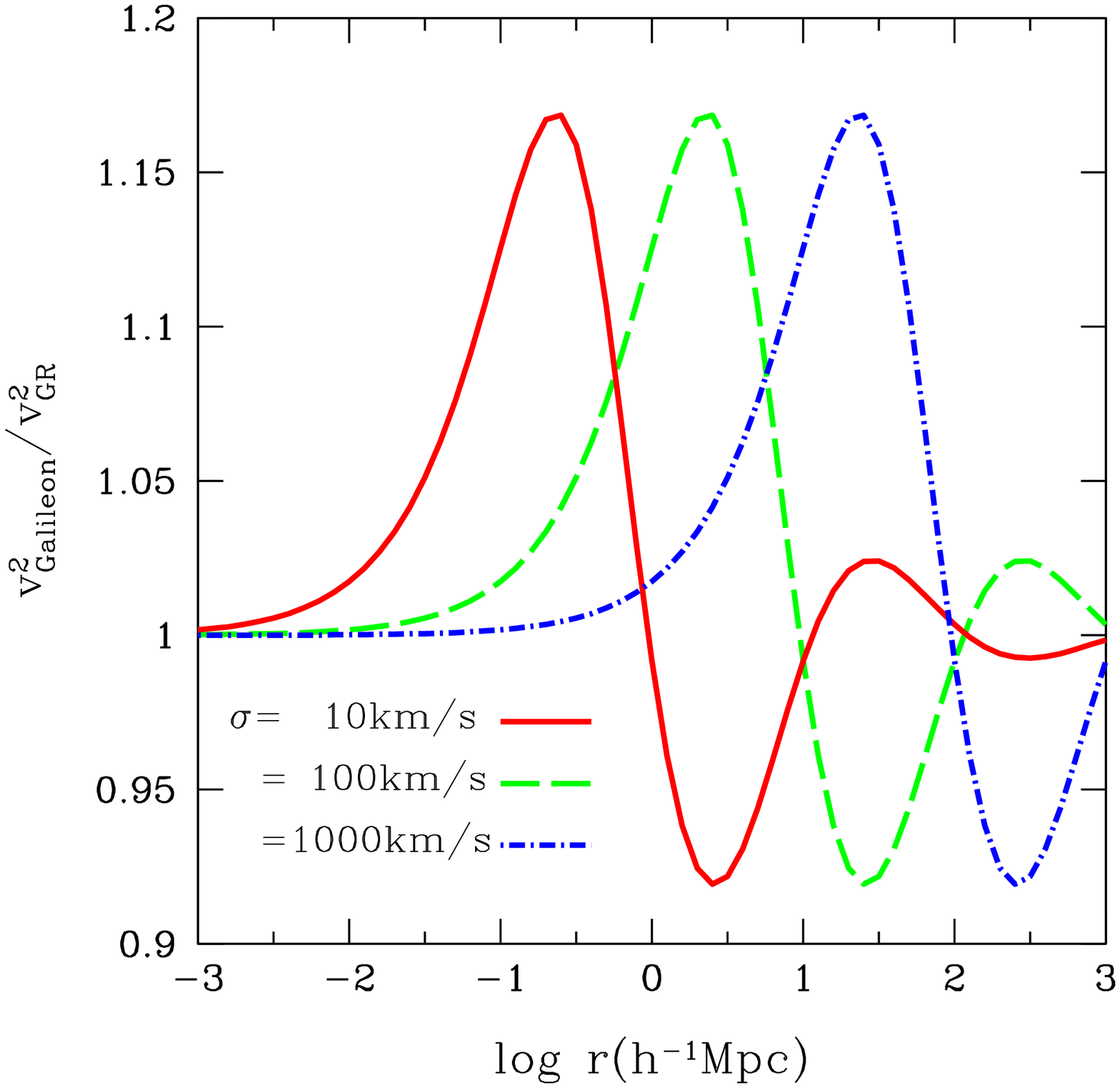}
      \end{center}
    \end{minipage}
  \end{tabular}
\caption{
The left panel is the ratio of the circular speed of a 
test particle for the SIS model in the sDGP model to
that in Newtonian gravity, as a function of radius $r(\hmpc)$.
Here, we adopted the velocity dispersion $\sigma=10{\rm km/s}$ 
(solid curve),  $\sigma=100{\rm km/s}$ (dashed curve), 
and $\sigma=1000{\rm km/s}$ (dot-dashed curve), respectively.
The right panel is the same as the left panel, but in the galileon model.
}
\label{fig:SIS_v}
\end{figure}

We next consider the circular speed of a test particle in a 
circular orbit at radius $r$, which is given by equating 
the gravitational force with the {\rm centrifugal force} 
\begin{eqnarray}
v^2(r)=r{d\Psi\over dr}.
\label{eq:vcirc}
\end{eqnarray}
Figure~\ref{fig:SIS_v} shows the ratio of circular speed
in the modified gravity model to that in Newtonian gravity, 
$v^2_{\rm MG}/v^2_{\rm GR}$, as a function of radius $r(\hmpc)$.  
Note that the circular speed in the SIS halo model satisfies 
$v^2_{\rm GR}=2\sigma^2$ in Newtonian gravity. 
The left (right) panel of Fig.~\ref{fig:SIS_v} shows the sDGP (galileon) model.
Each curve corresponds to the various velocity dispersion 
$\sigma=10{\rm km/s}$ (solid curve), 
$\sigma=100{\rm km/s}$ (dashed curve), 
and $\sigma=1000{\rm km/s}$ (dot-dashed curve), respectively.

Inside the Vainshtein radius $r\ll r_V$,  we have $v^2_{\rm MG}\simeq 
v^2_{\rm GR}$ due to the Vainshtein mechanism. However,  $v^2_{\rm MG}$
becomes to deviate from $v^2_{\rm GR}$ as $r$ becomes larger. 
The oscillatory feature appears at $r\sim r_V$. 
We suppose that the nonlinear structure of basic equations plays 
an important role in the oscillatory behavior around $r\sim r_V$. 
Finally, $v^2_{\rm MG}$ approaches $v^2_{\rm GR}$ again at 
large radii $r$ in both models. 
This behavior can be understood by considering the asymptotic 
solution of the density profile. Namely, 
in the limit $r\gg r_V$, using Eq.~(\ref{eq:Psilarger}), 
we obtain 
\begin{eqnarray}
v^2_{\rm MG}(r\gg r_V)\simeq{G_{\rm eff}M_{\rm MG}(r)\over r}\simeq v^2_{\rm GR}(r),
\label{eq:v_lr}
\end{eqnarray}
with the enclosed mass $M_{\rm MG}(r)=4\pi \int_0^rdr'r'^2 \rho_{\rm MG}(r')$.
Since the density of a halo in this limit is given by Eq.~(\ref{eq:rhoGali_lr}),
the effective gravitational constant is canceled out in Eq.~(\ref{eq:v_lr})
and the circular speed of a test particle in these modified gravity models
is equal to that of Newtonian gravity.
The characteristic behavior at $r\sim r_V$ 
appears well outside galaxies or galaxy clusters. 
Therefore, this effect of modification of gravity will be hard to be 
detected.

\begin{figure}[t]
      \begin{center}
\includegraphics[width=6.2cm,height=6.2cm,clip]{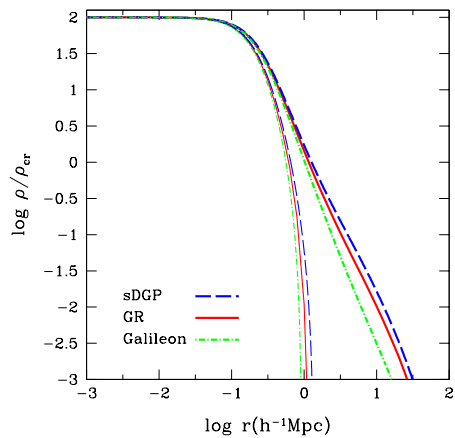}
      \end{center}
\caption{
The density profile of the King model 
as a function of radius $r(\hmpc)$.
The density is normalized by the critical density 
at present. The thick curves represent the density profile 
with the central potential $\Psi_0=-12\sigma^2$, 
while the thin curves represent the density profile with 
$\Psi_0=-3\sigma^2$.
The dashed, solid, and dot-dashed curve correspond to 
the sDGP model, Newtonian gravity, and the galileon model, respectively.
Here, the velocity dispersion is fixed as $\sigma = 100{\rm km/s}$.
}
  \label{fig:King_Dens}
\end{figure}

\subsection{King model}
The density profile of the King model 
can be obtained by solving Eqs. (\ref{eq:Poisson}), (\ref{eq:G_Field})
and (\ref{eq:DF_King}).
In Fig.~\ref{fig:King_Dens}, we show
the density profile of the King model normalized by the 
critical density at present, as a function of the radius, 
where the curves correspond to the sDGP model (dashed curve), 
Newtonian gravity (solid curve),
and galileon model (dot-dashed curve), respectively.
We adopted the values of the potential at the center, 
$\Psi_0=-12\sigma^2$ (thick curves) and $\Psi_0=-3\sigma^2$ 
(thin curves), respectively. 
Here, we set the velocity dispersion $\sigma = 100{\rm km/s}$.
In general, the halo density in the King model remains constant 
at small radii $r$, and becomes smaller as $r$ increases.

In the King model,  the density of a halo has a cut-off radius 
$r_t$, called the tidal radius, which is defined by $\Psi(r_t)=0$. 
From Eq.~(\ref{eq:DF_King}), we may set $\rho=0$
for $r\geq r_t$. 
Similar to the case of the SIS halo model, the density is enhanced (decreased)
in the sDGP (galileon) model in the outer region of halo in comparison with
that of Newtonian gravity, which 
explains the behavior of the density near the tidal radius 
in Fig.~\ref{fig:King_Dens}, depending on the models of gravity.

\begin{figure}[t]
  \begin{tabular}{cc}
    \begin{minipage}{0.5\textwidth}
      \begin{center}
\includegraphics[width=6.2cm,height=6.2cm,clip]{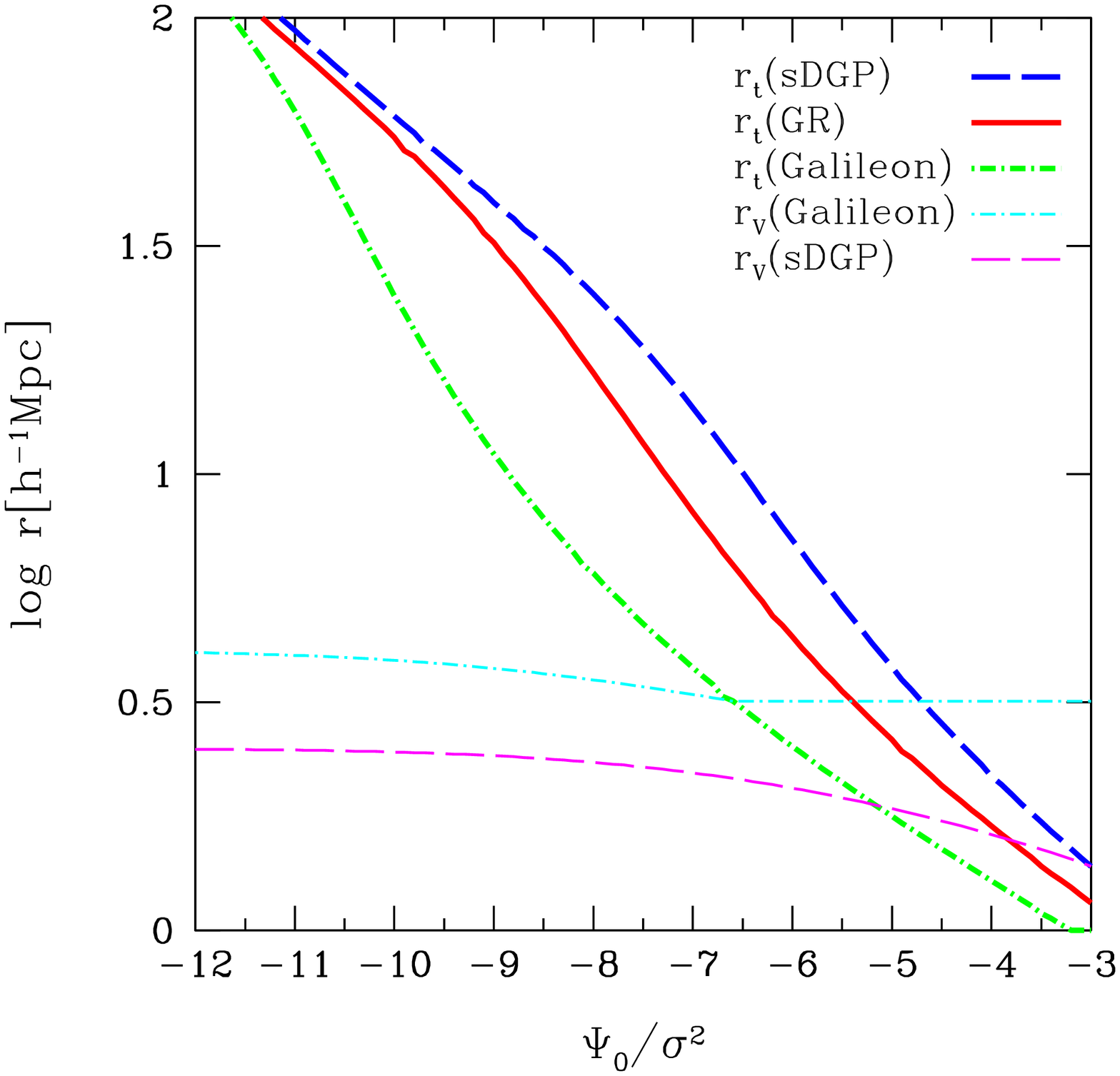}
      \end{center}
    \end{minipage}
    \begin{minipage}{0.5\textwidth}
      \begin{center}
\includegraphics[width=6.2cm,height=6.2cm,clip]{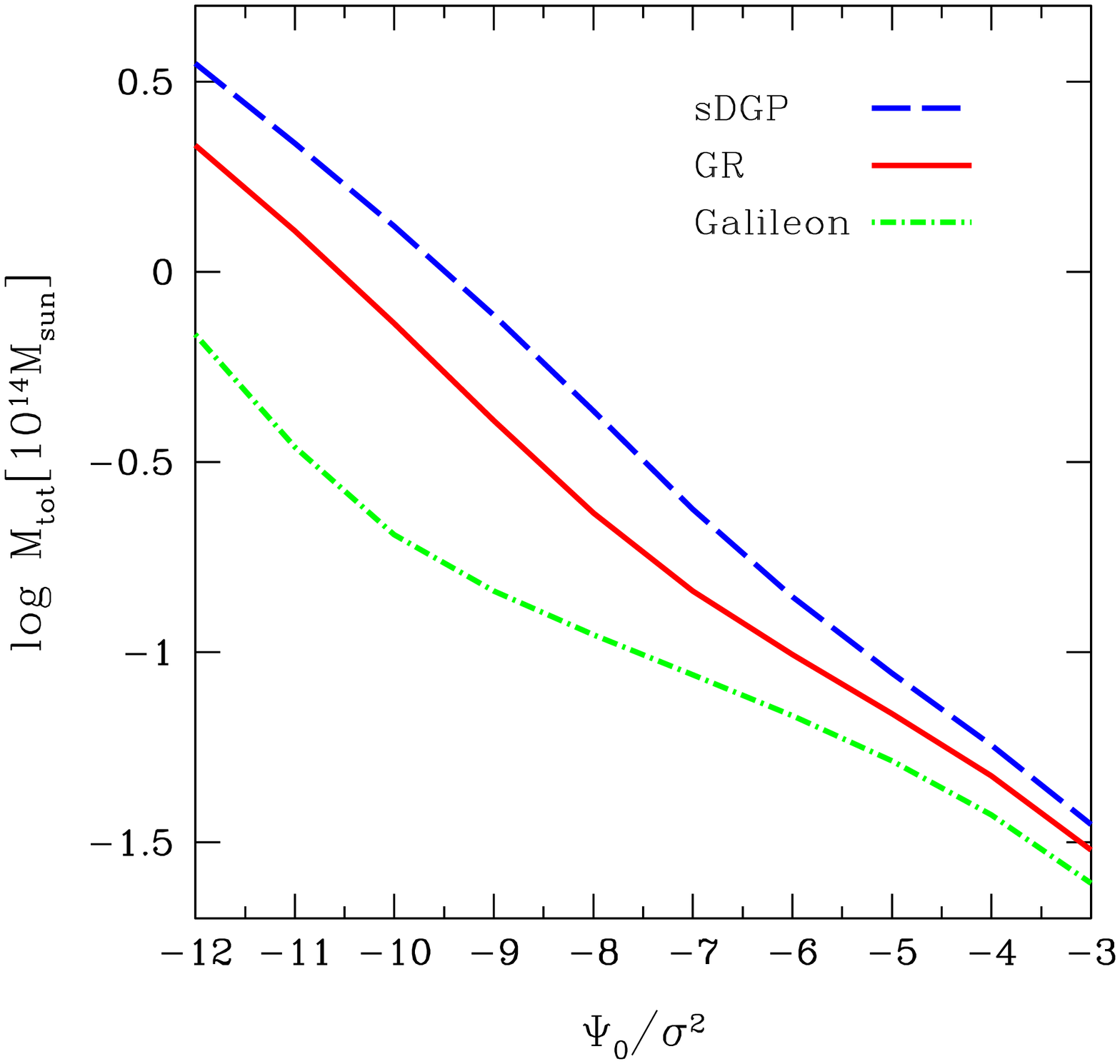}
      \end{center}
    \end{minipage}
  \end{tabular}
\caption{
The left panel plots the tidal radius (thick curves) and 
the Vainshtein radius (thin curves) of the King model as 
a function of $\Psi_0/\sigma^2$. 
Each thick curve is the tidal radius in the 
sDGP model (thick curve), Newtonian gravity (solid curve), 
and galileon model (dot-dashed curve), respectively.
The thin dashed (dot-dashed) curve is the 
Vainshtein radius in the sDGP (galileon) model, respectively.
The right panel plots the total mass $M_{\rm tot}$ as a function 
of $\Psi_0/\sigma^2$, for the sDGP (dashed curve), Newtonian gravity 
(solid curve), and galileon model (dot-dashed curve), respectively.
}
  \label{fig:King_V_Radius}
\end{figure}

We define the total mass of a halo by $M_{\rm tot}\equiv M(r_t)$.
In the left panel of Fig.~\ref{fig:King_V_Radius}, 
we plot the tidal radius $r_t$ as a function of central 
potential $\Psi_0/\sigma^2$ for the sDGP model (dashed curve),
Newtonian gravity (solid curve), 
and the galileon model (dot-dashed curve), respectively.
The right panel of Fig.~\ref{fig:King_V_Radius} plots 
the total mass as a function of the central potential, 
$\Psi_0/\sigma^2$, where   
each curve is the sDGP model (dashed curve),
Newtonian gravity (solid curve), 
and the galileon model (dot-dashed curve), respectively.
The total mass of the sDGP (galileon) model is larger 
(smaller) than that of Newtonian gravity.  
The difference of the total mass between the gravity models 
becomes larger as the absolute value of the central potential 
is larger.

In the left panel of Fig.~\ref{fig:King_V_Radius}, we also 
plot the Vainshtein radius $r_V$. 
The thin dashed curve and the thin dot-dashed curve 
represent the Vainshtein radius for the sDGP model and the 
galileon model, respectively.
In the case $r_t<r_V$,
the density profile at any radius is effectively described 
by Newtonian gravity, since all radii enclosing matter are 
inside the Vainshtein radius.
In the limit of small central potential $|\Psi_0|/\sigma^2$,
the tidal radius becomes very small, $r_t\ll r_V$.  
In this limit, therefore, the difference of the tidal radius 
between the gravity models becomes small. 
On the other hand, in the limit of large central potential 
$|\Psi_0|/\sigma^2$, the tidal radius becomes large, 
$r_t\gg r_V$. In the case $r_t>r_V$, the density profile 
in the outer region reflects the modification of gravity. 
Then, the difference of the tidal radius between 
the gravity models  remains in the limit 
of $|\Psi_0|/\sigma^2\gg1$ or $r_t\gg r_V$.
Let us now consider the question whether the gravity model can
be distinguishable or not. Even for the case $r_t>r_V$, 
the difference in the density profile appears only 
in the outer region of a halo.
Thus, the difference between those modified gravity 
theories and general relativity is small, and it will be 
difficult to be distinguished observationally.

\section{Discussion}
\label{discussion}

In Ref.~\refcite{dynamicalmass}, the author investigated 
the halos in the DGP modified gravity model with N-body 
simulations as well as analytic modeling of a halo similar 
to our investigation. Here, let us discuss the difference 
between the results in Ref.~\refcite{dynamicalmass} 
and ours. The author in Ref.~\refcite{dynamicalmass} 
has found that the square of the circular speed of a test 
particle is enhanced or decreased in outer region of a halo 
in proportion to $G_{\rm eff}$. On the contrary, our result 
in the SIS model is different, which the circular speed 
in the outer region finally approaches the value in 
the inner region of a halo. 
The difference comes from the treatment of the dark matter 
component. In our analysis, we adopted the explicit 
formula for the distribution function for the dark 
matter component, which is evidently the solution of 
the Boltzmann equation.
Thus, in our approach the same distribution function is 
assumed for the different gravity model, which results in 
the different density profiles, depending on the gravity models. 
 
\begin{figure}[t]
      \begin{center}
\includegraphics[width=6.2cm,height=6.2cm,clip]{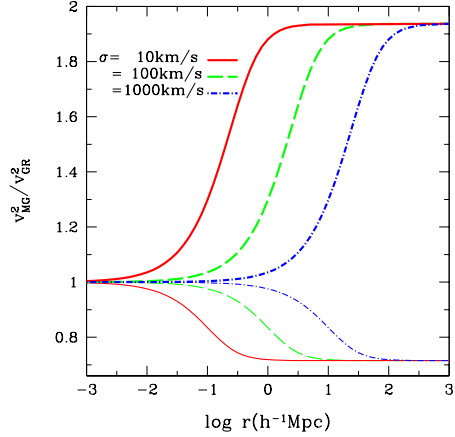}
      \end{center}
\caption{
The ratio of the circular speed of a test particle 
in the galileon model (thick curves) and the sDGP model (thin curves)
to that in Newtonian gravity for the SIS model, 
as a function of radius $r(h^{-1}{\rm Mpc})$, when the
density profile of a halo is fixed so as to be the SIS 
model, $\rho_{\rm MG}(r)=\sigma^{2}/2\pi Gr^{2}$. 
Here, we adopted the velocity dispersion $\sigma=10{\rm km/s}$ 
(solid curve),  $\sigma=100{\rm km/s}$ (dashed curve), 
and $\sigma=1000{\rm km/s}$ (dot-dashed curve), respectively.
For large $r$, this ratio approaches $G_{\rm eff}/G=1+\xi\zeta\simeq1.94$ 
for the galileon model, and $G_{\rm eff}/G=1+1/(3\beta)\simeq0.715$ for 
the sDGP model, respectively.}
  \label{fig:SIS_v2}
\end{figure}

On the other hand, the author of Ref.~\refcite{dynamicalmass} fixed 
the density profile of a halo independently of the gravity model, 
which has been suggested from N-body simulations. 
Following this method, the circular speed in a modified 
gravity model approaches that in Newtonian gravity 
multiplied by the factor $G_{\rm eff}/G$ in the limit of large radii. 
Fig.~\ref{fig:SIS_v2} shows the ratio of the circular speed 
of a test particle in the sDGP model and the galileon model 
to that in Newtonian gravity with fixing the density profile 
so as to be the SIS model, $\rho_{\rm MG}(r)=\sigma^{2}/2\pi Gr^{2}$. 
In Fig.~\ref{fig:SIS_v2}, we can see the similar feature of the 
circular speed in both the modified gravity models as that
expected from Ref.~\refcite{dynamicalmass}.

\begin{figure}[t]
      \begin{center}
\includegraphics[width=6.2cm,height=6.2cm,clip]{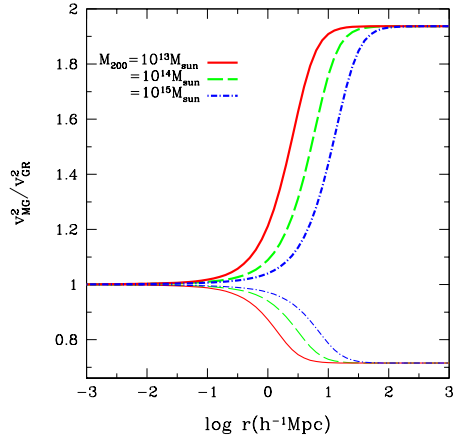}
      \end{center}
\caption{
The same as Fig.~\ref{fig:SIS_v2}, but for the NFW halo. 
Here, we adopted the virial mass $M_{200}=10^{13}\solM$ 
(solid curve), $10^{14}\solM$ (dashed curve), 
and $10^{15}\solM$ (dot-dashed curve), respectively.
The asymptotic value of $G_{\rm eff}/G$ at large $r$ is 
same as that in Fig.~\ref{fig:SIS_v2}. }
  \label{fig:NFW_v2}
\end{figure}

Following Ref.~\refcite{dynamicalmass}, we next consider an 
Navarro-Frenk-White\cite{NFW} (NFW) halo with mass $M_{200}$, 
defined as the mass contained with a radius $R_{200}$ so that 
the average density within $R_{200}$ is $200 \bar \rho $,
where $\bar \rho$ is the background density, and we used the 
concentration relation\cite{Bullock}, 
$c=9\times(M/[3.2\times 10^{12}\solM/h])^{-0.13}$. 
In this case, the virial radius is written 
$R_{200}\simeq1.0\times(M_{200}/10^{14}\solM)^{1/3}$~$h^{-1}$Mpc, 
and the concentration is 
$c\simeq6.0\times(M_{200}/10^{14}\solM)^{-0.13}$ (see, e.g., 
Ref.~\refcite{dynamicalmass} for details). 
Fig.~\ref{fig:NFW_v2} shows the ratio of the circular speed
in the modified gravity models to that in  
Newtonian gravity. Similar feature as that in 
Fig.~\ref{fig:SIS_v2} can be seen for the NFW halo. 
In Fig.~\ref{fig:NFW_v2}, we choose the mass $M_{200}=10^{13}\solM$ (solid curve),
~$10^{14}\solM$ (dashed curve),~$10^{15}\solM$ (dot-dashed curve), 
which correspond to clusters. 
The effect of the modified gravity is significant in the outer 
region of a halo, but it is suppressed in the inner region 
because of the Vainshtein mechanism.
For the NFW halo, the Vainshtein radius is estimated as
\begin{eqnarray}
r_V\simeq5.6\left({M_{200}\over 10^{14}\solM}\right)^{1/3}\hmpc,
\end{eqnarray}
for the sDGP model, and 
\begin{eqnarray}
r_V\simeq11\left({M_{200}\over 10^{14}\solM}\right)^{1/3}\hmpc,
\end{eqnarray}
for the galileon model, respectively, from Eq.~(\ref{eq:V_Radius}). 
The Vainshtein radius is large, but the Vainshtein mechanism 
does not completely hide the effect of the modification of 
gravity in the cluster. One can read that the circular speed
is enhanced or decreased in comparison with Newtonian 
halo, e.g., at $10$\% level at the radius of a few  
$h^{-1}{\rm Mpc}$ for $M_{200}=10^{14}\solM$.
If a good tracer of circular speed is available, 
there might be a possible chance to test the 
modified gravity effect in a halo, as discussed 
in Ref.~\refcite{dynamicalmass}.

\section{Summary and conclusions}
\label{conclusion}
In this paper, we have investigated the structure of dark matter halos 
in the sDGP model and the galileon model by using the static and 
spherically symmetric solutions of the collisionless Boltzmann equation, 
which reduce to the SIS model and the King model in the limit of 
Newtonian gravity. 
We have obtained the solution of a halo in these modified gravity 
theories in a numerical manner.  
The density of the halo is the same as that in Newtonian gravity 
due to the Vainshtein mechanism well inside the Vainshtein radius 
$r\ll r_V$. We have found 
that the density in the sDGP (the galileon) model becomes 
larger (smaller) in the outer region of a halo, in comparison with that in
Newtonian gravity, which comes from the suppression 
(enhancement) of the effective gravitational constant in the 
sDGP (galileon) model, respectively.
In the SIS model, the density of the halo at large radii
$r\gg r_V$ approaches the value of Newtonian gravity 
multiplied by a factor $G/G_{\rm eff}$. 
The circular speed is also modified around $r\sim r_V$. 
However, the Vainshtein radius is $r_V\simeq 1\sim10{\rm Mpc}$ 
for a typical galaxy 
and $r_V\simeq 10\sim100{\rm Mpc}$ for a typical galaxy cluster. 
Therefore, it will be difficult to distinguish between 
these modified gravity theories by a measurement of a 
halo because the Vainshtein radius is large enough.  

However, the above results rely on the assumption that
the distribution function of the dark matter follows the simple
explicit formula. For a comparison with the previous approach 
in Ref.~\refcite{dynamicalmass}, we investigated the effect of 
the modification of gravity with fixing the halo density profile
independently of the gravity model. 
In this case, the circular speed is enhanced or decreased 
in the outer region of a halo depending on the gravity model. 
In the case adopting the NFW profile as a model of a cluster halo, 
the Vainshtein radius is large.  However, the Vainshtein mechanism 
does not completely hide the effect of the modification of gravity 
in the cluster. This might provide a possible chance that precise 
measurement of halo could be a probe of the modified gravity, 
as discussed in Ref.~\refcite{dynamicalmass}. 

\vspace{3mm}
\section*{Acknowledgments}
We thank anonymous referee for useful comments which helped
improve the original manuscript. 
This work was supported by Japan Society for Promotion
of Science (JSPS) Grants-in-Aid
for Scientific Research (Nos.~21540270,~21244033).
This work was also supported by JSPS 
Core-to-Core Program ``International Research 
Network for Dark Energy''.
TN and RA acknowledge support by a research assistant program
of Hiroshima University. 
This work was supported in part by a Grant-in-Aid for JSPS Fellows (TN).


\begin{thebibliography}{0}
\bibitem{Riess}
  A. G. Riess et al., Astron. J. {\bf 116} (1998) 1009   
\bibitem{Perlmutter}
  S. Perlmutter et al., Astropys. J. {\bf 517} (1999) 565  
\bibitem{Maartens}
  R. Durrer, R. Maartens, arXiv:0811.4132 
\bibitem{Jain10}
  B. Jain and J. Khoury, Annals of Physics {\bf 325} (2010) 1479
\bibitem{Tsujikawa}
 S. Tsujikawa, Lect.~Notes Phys {\bf 800} (2010) 99
\bibitem{DE06} E.~J. Copeland, M. Sami and S. Tsujikawa, Int. J. Mod. Phys. 
{\bf D15} (2006) 1753
\bibitem{AmeTsujiDE}
  L. Amendla and S. Tsujikawa,  {\it Dark Energy: Theory and Observations} (Cambridge University Press, 2010)
\bibitem{DGP}
  G. R. Dvali, G. Gabadadze and M. Porrati, Phys. Lett. B {\bf 485} (2000) 
 208 
\bibitem{DGP2}
  C. Deffayet, Phys. Lett. B {\bf 502} (2001) 199 
\bibitem{KoyamaMaartens}
  K. Koyama and R. Maartens, JCAP {\bf 01} (2006) 016 
\bibitem{Nicolis04}
  A. Nicolis and R. Rattazzi, JHEP {\bf 0406} (2004) 059 
\bibitem{Gorbunov06}
  D. Goubnov, K. Koyama and S. Sibiryakov, Phys. Rev. D {\bf 73} 
 (2006) 044016 
\bibitem{Fairbairn}
  M. Fairbairn and A. Goobaar, Phys. Lett. B {\bf 642} (2006) 432 
\bibitem{Maartens06}
  R. Maartens and E. Majerotto, Phys. Rev. D {\bf 74} (2006) 023004 
\bibitem{Song07}
  Y. S. Song, I. Sawicki and W. Hu, Phys. Rev. D {\bf 75} (2007) 064003 
\bibitem{DGP3}
  C. Deffayet, G. R. Dvali and G. Gabadadze, Phys. Rev. {\bf D65} (2002) 044023 
\bibitem{Luty03}
  M. A. Luty, M. Porrati and R. Rattazzi, JHEP {\bf 0309} (2003) 029 
\bibitem{GALMG}
  A. Nicolis, R. Rattazzi and E. Trincherini, 
  Phys. Rev. D {\bf 79} (2009) 064036 
\bibitem{CG}
  C. Deffayet, G. Esposito-Farese and A. Vikman, 
  Phys. Rev. D {\bf 79} (2009) 084003 
\bibitem{GC}
  N. Chow and J. Khoury, Phys. Rev. D {\bf 80} (2009) 024037 
\bibitem{SAUGC}
  F. P. Silva and K. Koyama, Phys. Rev. D {\bf 80} (2009) 121301 
\bibitem{ELCP}
  T. Kobayashi, H. Tashiro and D. Suzuki, Phys. Rev. D {\bf 81} (2010) 063513 
\bibitem{CEGH}
  T. Kobayashi, Phys. Rev. D {\bf 81} 103533 (2010)
\bibitem{Deffayet} 
  C. Deffayet, O. Pujolas, I. Sawicki and A. Vikman, JCAP {\bf 10} (2010) 026 
\bibitem{GInflation}
  T. Kobayashi, M. Yamaguchi and J. Yokoyama, Phys. Rev. Lett. {\bf 105} (2010) 231302 
\bibitem{DeFelice10}
  A. De Felice and S. Tsujikawa, arXiv:1008.4236
\bibitem{CCGF}
  A. De Felice and S. Tsujikawa, Phys. Rev. Lett. {\bf 105} (2010) 111301 
\bibitem{OCG} 
  S. Nesseris, A. De Felice and S. Tsujikawa, Phys. Rev. D {\bf 82} (2010) 124054 
\bibitem{Kimura}
  R. Kimura and K. Yamamoto, JCAP {\bf 04} (2011) 025 
\bibitem{Vainshtein}
  A. I. Vainshtein, Phys. Lett. B {\bf 39} (1972) 393 
\bibitem{dynamicalmass}
  F. Schmidt, Phys. Rev. D {\bf 81} (2010) 103002 
\bibitem{Wyman}
  M. Wyman, Phys. Rev. Lett. {\bf 106} (2011) 201102
\bibitem{Konno}
  K. Konno et al., Phys. Rev. D {\bf 78} (2008) 024037
\bibitem{Gergely}
 L. A. Gergely et al., arXiv:1105.0159
\bibitem{Burikham}
 P. Burikham and S. Panpanich, arXiv:1103.1198 
\bibitem{Koyama07}
  K. Koyama and F. P. Silva, Phys. Rev. D {\bf 75} (2007) 084040  
\bibitem{Binney}
  J. Binney and S. Tremain, {\it Galactic Dynamics} (Princeton University Press, 2008) 
\bibitem{NFW}
  J. F. Navarro, C. S. Frenk and 
  S. D. M. White,  Astrophys. J. {\bf 490} (1997)
493 
\bibitem{Bullock}
  J. S. Bullock, et al., Mon. Not. R. Astron. Soc. {\bf 321} (2001) 559 
\end{thebibliography}
\end{document}